# Electrically and magnetically induced optical rotation in $Pb_5Ge_3O_{11}$:Cr crystals at the Phase Transition.

## 2. Faraday effect in pure and Cr-doped lead germanate in the presence of electric field and spontaneous polarization

Adamenko D. [1], Klymiv I. [1], Duda V.M. [2], Vlokh R. [1] and Vlokh O. [1]

[1]Institute of Physical Optics, 23 Dragomanov St., 79005 Lviv, Ukraine, e-mail: vlokh@ifo.lviv.ua

[2]Dnipropetrovsk National University, 13 Naukova St., Dnipropetrovsk, Ukraine

**Abstract**

This work presents the results for the Faraday rotation in pure and Cr-doped lead germanate crystals studied in the course of proper ferroelectric phase transition. We show that the increment of the Faraday rotation appearing at the phase transition is caused by a combined magneto-electrooptic effect induced by spontaneous polarization. It is proportional to the square of spontaneous polarization. The phenomenon revealed by us corresponds to combined effects of crystal optics, which appear due to common action of different fields.

## Introduction

The Faraday effect is a well-known phenomenon consisting in rotation of polarization plane of light ($\rho_k$) under the action of magnetic field ($H_l$). The authors [1-3] have experimentally shown that a rotation, which is additional to the Faraday one, can appear when a spontaneous polarization or biasing electric field are present in ferroelectric crystals. The corresponding rotation angle is proportional to both the magnetic field and the electric field (or the electric polarization). From the viewpoint of symmetry, this effect should be described by the same symmetry as the Faraday effect [4]. A possibility for existence of such the effect has once been mentioned in the monograph [5]. According to [5], the effect might have been described by antisymmetric fourth-rank tensor $A_{ijlm} = -A_{jilm}$ ($\Delta\varepsilon_{ij}(\omega, k, E^{(0)}, H^{(0)}) = A_{ijlm}(\omega) E_l^{(0)} H_m^{(0)}$). It should therefore manifest itself as an optical rotation appearing under the conditions of common action of electric and magnetic fields. Following [5], the optical-frequency dielectric permittivity for magnetically non-ordered crystals in the presence of both magnetic and electric fields may be presented in the form

$$\varepsilon_{ij}(\omega,k,E^{(0)},H^{(0)}) = \varepsilon_{ij}^{0} + A_{ijl}(\omega)E_l^{(0)} + A_{ijm}(\omega)H_m^{(0)} + A_{ijlm}(\omega)E_l^{(0)}H_m^{(0)} + A_{ijlmn}(\omega)E_l^{(0)}H_m^{(0)}k_n , \quad (1)$$

where $\varepsilon_{ij}^{0}$ means the unperturbed dielectric permittivity, $A_{ijl}(\omega) = A_{jil}(\omega)$ a tensor describing the Pockels effect, $A_{ijm}(\omega) = -A_{jim}(\omega)$ the same for the Faraday effect and $A_{ijlmn} = A_{jilmn}$ a fifth-rank pseudo-tensor governing the changes in the refractive indices imposed by common action of electric and magnetic fields, with accounting for the spatial dispersion effects. Notice that the last term in Eq. (1) cannot be responsible for the additional optical rotation observed in the alum crystals after application of electric and magnetic fields, as mentioned in [6].

Let us consider a polar Faraday tensor $\alpha_{kl}$ appearing in the relation

$$\rho_k = \frac{\pi}{\lambda n}(\alpha_{kl}H_l + \delta_{klm}H_lE_m + \Theta_{klmn}H_lE_mE_n) = \frac{\pi}{\lambda n}(\alpha_{kl} + \delta_{klm}E_m + \Theta_{klmn}E_mE_n)H_l , \quad (2)$$

where $\lambda$ is the light wavelength and $n$ the refractive index. It follows from Eq. (1) that the tensor $\alpha_{kl}$ can be changed under the action of electric field on the optical medium:

$$\Delta\alpha_{kl} = \delta_{klm}E_m + \Theta_{klmn}E_mE_n , \quad (3)$$

where $E_m$ and $E_n$ are the electric field components and $\delta_{klm}$ and $\Theta_{klmn}$ represent third- and fourth-rank polar tensors, respectively. The phenomenon mentioned above might be also treated as a combined magneto-electrooptic rotation, which appears due to common action of the electric and magnetic fields and manifests itself as a rotation additional to the Faraday one. From this point of view, the conclusion [7] about a necessity of independence of the additional rotation of the optical path in crystal seems to be unclear. Obviously, this effect should be much smaller than the Faraday one and so hard to detect experimentally. Nevertheless, this combined effect has been revealed in [3] as an increment of the Faraday rotation induced by spontaneous polarization appearing in the course of proper ferroelectric phase transition in lead germanate (LG) crystals. However, the authors [1,3] have claimed observation of the Faraday effect increments both linear and quadratic in the spontaneous polarization, while the linear one should be forbidden by the symmetry of paraelectric (PE) phase of LG.

In the first part of our studies (see [8]) we have demonstrated that $Pb_5Ge_3O_{11}$:Cr crystals manifest a proper, second-order ferroelectric phase transition at $T_c = 454\,\mathrm{K}$, at which the changes in the optical properties such as the optical activity and the electrogyration effect are well described by spontaneous polarization as the order parameter. This implies that the changes in the optical rotatory power below the phase transition point are fully described by the electrogyration effect induced by spontaneous polarization, while the electrogyration coefficient follows the Curie-Weiss law in the vicinity of $T_c$. In this case the behaviour of the other constitutive coefficients for $Pb_5Ge_3O_{11}$:Cr crystals should obey regularities, which are typical for the proper second-order phase transitions. For example, the temperature changes of the Faraday coefficients $\Delta\alpha_{kl}$ below the phase transition temperature should be proportional to the square of spontaneous polarization ${}^sP_m{}^sP_n$:

$$\Delta\alpha_{kl} = \tilde{\Theta}_{klmn}{}^sP_m{}^sP_n\,, \qquad (4)$$

where $\tilde{\Theta}_{ijkl}$ is a fourth-rank polar tensor. Since the magnetooptic rotation $\rho_k$ is described by Eq. (2), the change in the optical rotatory power induced by both magnetic field and spontaneous polarization can be presented as a result of their combined effect:

$$\rho_k = \frac{\pi}{\lambda n}\tilde{\Theta}_{klmn}H_l{}^sP_m{}^sP_n\,. \qquad (5)$$

or

$$\rho_k = \frac{\pi}{\lambda n}\Theta_{klmn}H_lE_mE_n\,, \qquad (6)$$

in terms of the electric field. Then the coefficients $\tilde{\Theta}_{klmn}$ could be determined from the increments of the Faraday coefficients at the proper ferroelectric phase transition and compared with the coefficients obtained experimentally (see, e.g., [2]).

The present paper is devoted to elucidation of dependence of the Faraday rotation increment on the temperature and spontaneous polarization. Furthermore, we will clarify dependence of the combined electro-magnetooptic rotation on the optical path and estimate the tensor component describing the corresponding effect. In our experiments we have used both the pure LG and $Pb_5Ge_3O_{11}$ crystals doped with $1.20 \pm 0.08$ weight % of Cr ions.

**Experimental**

Since $Pb_5Ge_3O_{11}$ and $Pb_5Ge_3O_{11}$:Cr crystals possess the second-order phase transition with the change of point symmetry group $\bar{6} \leftrightarrow 3$ [8], the Faraday tensor in the both structural phases has the form

$$\alpha_{kl} = \begin{array}{c|ccc} & H_1 & H_2 & H_3 \\ \hline \rho_1 & \alpha_{11} & 0 & 0 \\ \rho_2 & 0 & \alpha_{11} & 0 \\ \rho_3 & 0 & 0 & \alpha_{33} \end{array}\,. \qquad (7)$$

The Faraday rotation observed in our experimental geometry ($k \| H \| z$) is defined by the formula

$$\rho_3 = \frac{\pi}{\lambda n_o} \alpha_{33} H_3 , \qquad (8)$$

whereas the total optical rotation induced by the magnetic field and the electric field (or the spontaneous polarization) may be written as

$$\rho_3 = \frac{\pi}{\lambda n} (\alpha_{33} + \Theta_{3333} E_3 E_3) H_3 , \qquad (9)$$

or

$$\rho_3 = \frac{\pi}{\lambda n} (\alpha_{33} + \tilde{\Theta}_{3333} {}^{s}P_3 {}^{s}P_3) H_3 . \qquad (10)$$

The experimental set-up is presented in Fig. 1. The sample was placed into optical heating stage allowing temperature stabilization with the accuracy of 0.1 K. The electric field was applied along the direction $< 001 >$ in crystals, using glass electrodes coated by conducting tin oxide layer. The magnetic field was applied along the same direction with the aid of electromagnet. The absolute values of Faraday coefficients were determined with the error  12%, while the relative changes in the Faraday rotation were measured more precisely (the error did not exceed 2%). Note that some extra Faraday rotation ($\sim 10\%$) appeared in the optical windows and transparent glass electrodes, because these optical elements were subjected to the magnetic field, too. This additional rotation was also taken into consideration when calculating the Faraday coefficient.

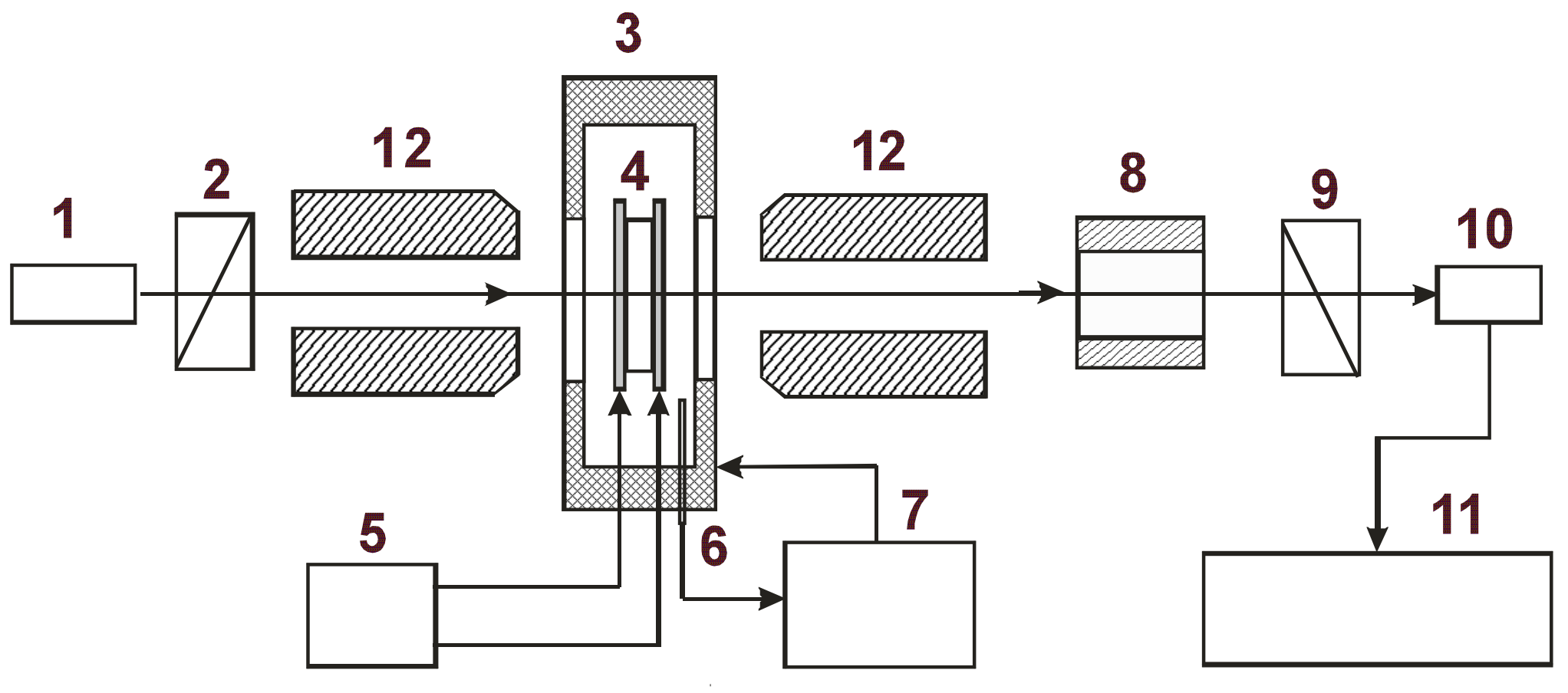

Fig. 1. Experimental set-up: 1 – He-Ne laser, 2, 9 – polarizers with rotation stages, 3 – furnace, 4 – sample with transparent electrodes, 5 – high-voltage source, 6 – thermocouple, 7 – temperature controller, 8 – Faraday cell, 10 – photomultiplier, 11 – oscilloscope, 12 – electromagnet.

## Results and discussion

The temperature dependences of the Faraday coefficients for $Pb_5Ge_3O_{11}$ and $Pb_5Ge_3O_{11}:Cr^{3+}$ crystals are presented in Fig. 2. As one can see, the Faraday coefficients change their behaviour in the vicinity of phase transition. The temperature dependence in the PE phase has been fitted by a power function $\Delta\alpha_{33} = AT + BT^2$ and then approximated into the range of ferroelectric phase. The difference between the approximating curve and the temperature dependence experimentally obtained for the ferroelectric phase corresponds to the increment of Faraday coefficient caused by spontaneous polarization.

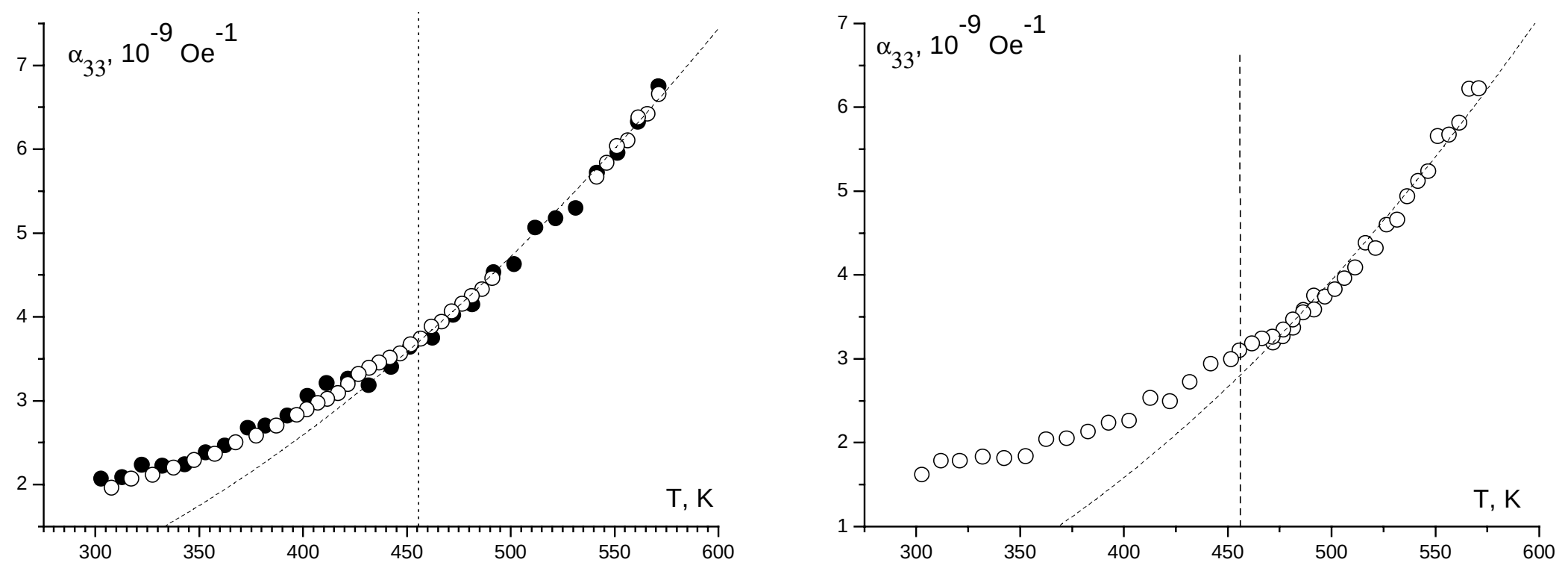

a b

Fig. 2. Temperature dependences of the Faraday coefficients for $Pb_5Ge_3O_{11}$ (a) and $Pb_5Ge_3O_{11}:Cr^{3+}$ (b) crystals at $\lambda = 632.8\,\mathrm{nm}$. Dashed line represents extrapolation of temperature curve for the Faraday coefficients from the PE phase into the ferroelectric one. Open and full circles in Fig. 2a correspond to the samples thicknesses $d = 8.92\,\mathrm{mm}$ and $d = 5.52\,\mathrm{mm}$, respectively.

The fitting coefficients are equal to $A = -0.00536 \times 10^{-9} (Oe\,K)^{-1}$ and $B = 2.96 \times 10^{-14} (OeK^2)^{-1}$ for the pure LG crystals, while for the Cr-doped crystals the corresponding values are $A = -0.01 \times 10^{-9} (Oe\,K)^{-1}$ and $B = 3.92 \times 10^{-14} (OeK^2)^{-1}$. The dependences of increments of the Faraday coefficients on the spontaneous polarization are presented in Fig. 3 (notice that we have used the spontaneous polarization data for the pure LG for the both crystals studied here – see [9]). As seen from Fig. 3, these dependences are well fitted by a quadratic function (see Eq. (4)).

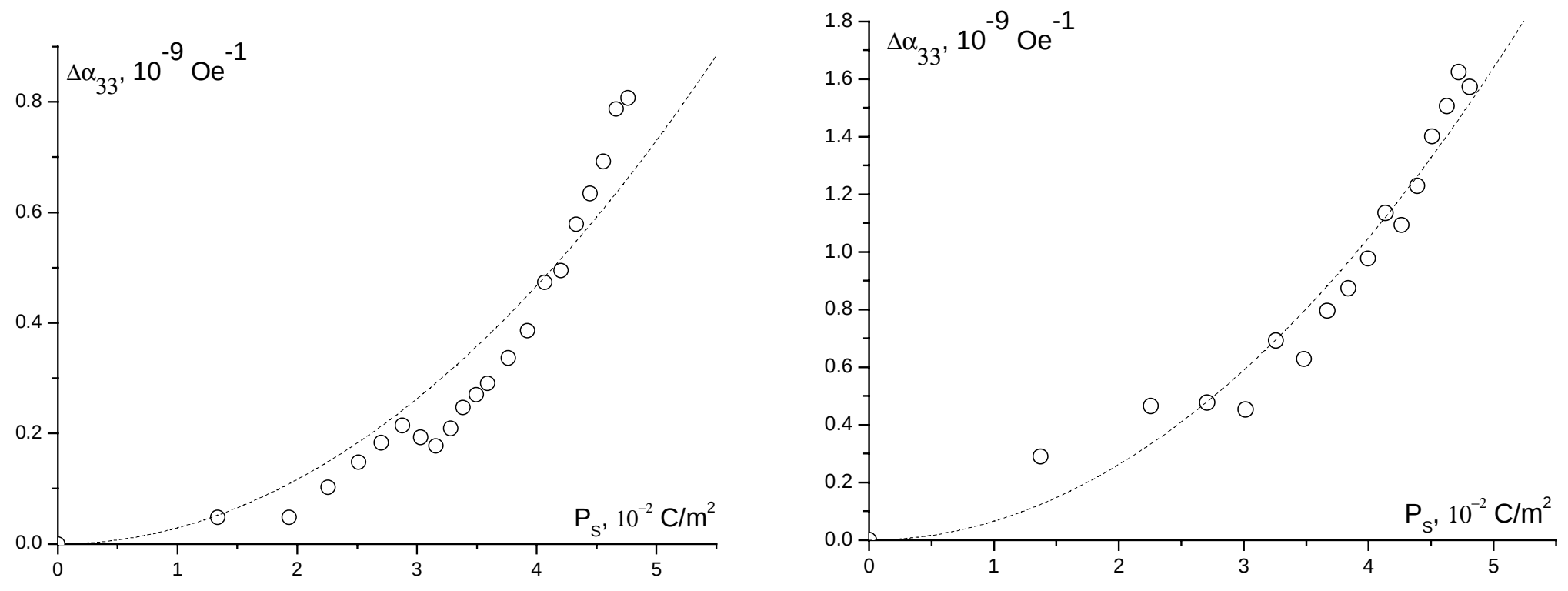

a b

Figure 3. Dependences of increments of the Faraday coefficients on the spontaneous polarization for $Pb_5Ge_3O_{11}$ (a) and $Pb_5Ge_3O_{11}:Cr^{3+}$ (b) crystals at $\lambda = 632.8\,\text{nm}$. Open circles correspond to the experimental data and dashed curves to the fitting by quadratic function.

On the basis of Eq. (4) we have calculated the coefficients of combined electro-magnetooptic effect for the PE phase. The coefficient is equal to $\tilde{\Theta}_{3333} = 2.9\times10^{-7}\,\text{m}^4/(\text{Oe}\times\text{C}^2)$ for the pure LG crystals, while for the Cr-doped crystals it is larger, $\tilde{\Theta}_{3333} = 6.6\times10^{-7}\,\text{m}^4/(\text{Oe}\times\text{C}^2)$. Using the relation

$$\Theta_{3333} = \tilde{\Theta}_{3333}\varepsilon_0^2(\varepsilon_{33} - 1)^2 \simeq \tilde{\Theta}_{3333}\varepsilon_0^2\varepsilon_{33}^2, \qquad (11)$$

with $\varepsilon_{33}$ being the dielectric permittivity of the pure LG crystals [9] and $\varepsilon_0$ the free-space dielectric permittivity, we have calculated temperature dependences of the coefficients $\Theta_{3333}$ in the PE phase for the both crystals (see Fig. 4) and the specific optical rotation referred to the electric and magnetic fields $E = 10^6\,\text{V/m}$ and $H = 13.3\,\text{kOe}$, respectively (notice that just these values of the electric and magnetic fields have been used in [2] in the studies of magneto-electrooptic rotation). We remind here that, according to [2], magneto-electrooptic rotation in the PE phase, being proportional to the square of electric field, has been nearly zero for $Pb_5Ge_3O_{11}:Gd^{3+}$ and equal to $0.3\,\text{deg/cm}$ for $Pb_5Ge_3O_{11}:Nd^{3+}$ crystals in the vicinity of $\text{T}_\text{c}$. The effect bilinear in the both fields ($\rho \sim EH$) has been revealed in the ferroelectric phase close to $T_c$. As one can see (Fig. 4), the magneto-electrooptic rotation in the PE phase exceeds the experimental error and so can be detected experimentally.

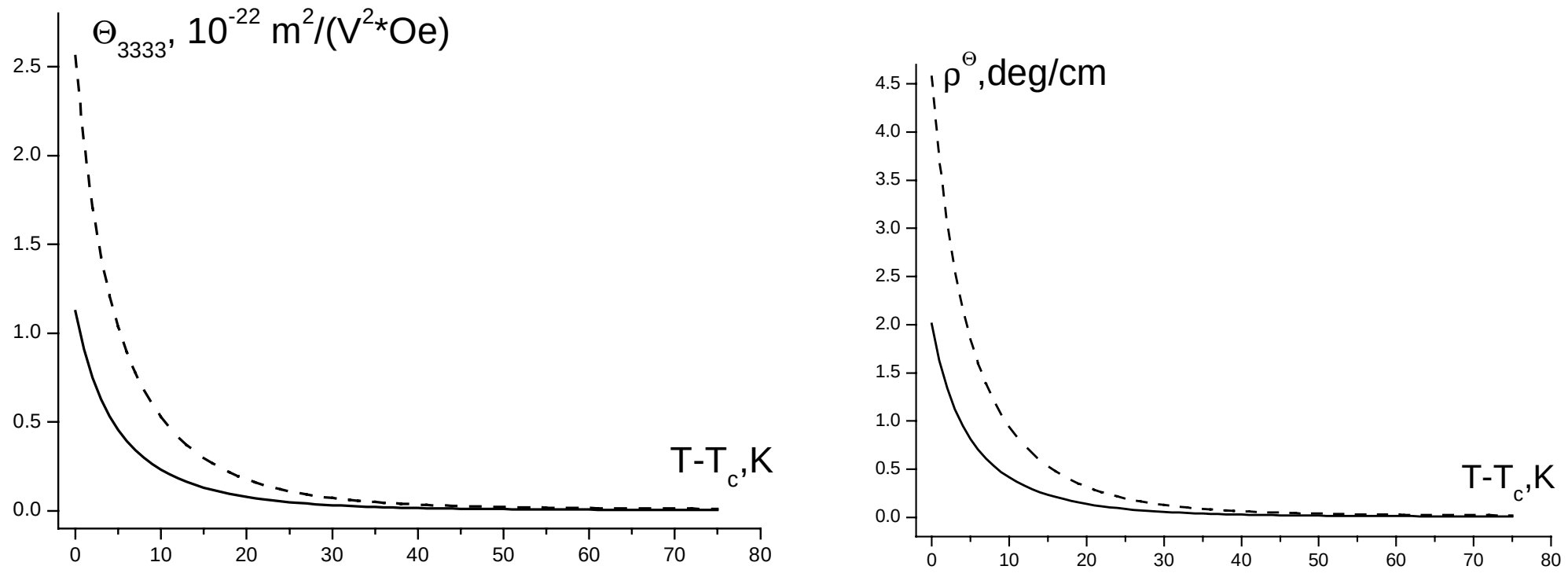

a b

Figure 4. Calculated temperature dependences of the coefficient $\Theta_{3333}$ (a) and the specific optical rotation (b) induced by common action of the electric ($E = 10^6$ V/m) and magnetic ($H = 13.3$ kOe) fields for $Pb_5Ge_3O_{11}$ (solid curve) and $Pb_5Ge_3O_{11}$:$Cr^{3+}$ (dashed curve) crystals. The light wavelength is $\lambda = 632.8$ nm.

The electric and magnetic fields used in our present experiments ($E = 0.3\times10^6$ V/m and $H = 3$ kOe at $d = 5.74$ mm) are smaller than those dealt with in the work [2]. According to the estimations, the optical rotation for such the field and sample thickness values should be about $\sim 3'$, which could hardly be carefully measured in our experiment. In fact, we have not observed any additional rotation induced by the common action of electric and magnetic fields in the overall temperature interval under study. Ill-defined changes in the Faraday rotation in the electric field have been detected only in the vicinity of $T_c$. Probably, they are caused by a shift of phase transition temperature induced by the biasing field.

In the present study we have also measured temperature dependence of the Faraday coefficient for the pure LG crystals, using the samples with different thicknesses (see Fig. 2a). It has been found that the increment of the Faraday rotation caused by spontaneous polarization differs for the samples with different thicknesses, thus being evidently dependent on the optical path. As a consequence, the prediction of the work [7] concerning the independence of magneto-polarizational (or magneto-electric) phase difference on the optical path is questionable.

**Conclusions**

In the present work we have revealed magneto-electrooptic rotation induced by the external magnetic field and the spontaneous polarization in $Pb_5Ge_3O_{11}$ and $Pb_5Ge_3O_{11}$:$Cr^{3+}$ crystals. However, we have not detected the corresponding effect induced by the biasing electric field, because it is too small to be detected with the electric fields at our disposal. The value of magneto-electrooptic coefficient estimated for the PE phase is equal to $\tilde{\Theta}_{3333} = 2.9\times10^{-7}\,\mathrm{m}^4/(\mathrm{Oe}\times\mathrm{C}^2)$ in case of $Pb_5Ge_3O_{11}$ and $\tilde{\Theta}_{3333} = 6.6\times10^{-7}\,\mathrm{m}^4/(\mathrm{Oe}\times\mathrm{C}^2)$ for $Pb_5Ge_3O_{11}:Cr^{3+}$ crystals. It has been shown that the phase difference responsible for the magneto-electrooptic rotation and induced by the common action of electric and magnetic fields depends on the optical path.

**Acknowledgement**

The authors acknowledge financial support of this study from the Ministry of Education and Science of Ukraine (the Project N0106U000616).